\documentclass[twoside,twocolumn,9pt]{article}
\usepackage{extsizes}
\usepackage[super,sort&compress,comma]{natbib} 
\usepackage[version=3]{mhchem}
\usepackage[left=1.5cm, right=1.5cm, top=1.785cm, bottom=2.0cm]{geometry}
\usepackage{balance}
\usepackage{times,mathptmx}
\usepackage{sectsty}
\usepackage{graphicx} 
\usepackage{lastpage}
\usepackage[format=plain,justification=justified,singlelinecheck=false,font={stretch=1.125,small,sf},labelfont=bf,labelsep=space]{caption}
\usepackage{float}
\usepackage{fancyhdr}
\usepackage{fnpos}
\usepackage[english]{babel}
\usepackage{array}
\usepackage{droidsans}
\usepackage{charter}
\usepackage[T1]{fontenc}
\usepackage[usenames,dvipsnames]{xcolor}
\usepackage{setspace}
\usepackage[compact]{titlesec}

\usepackage{epstopdf}

\definecolor{cream}{RGB}{222,217,201}

\newcommand{\nvec}{\mathbf{n}}

\newcommand{\rvec}{\mathbf{r}}

\newcommand{\Qvec}{\mathbf{Q}}
\newcommand{\Pvec}{\mathbf{P}}

\DeclareMathOperator{\Tr}{Tr}

\begin{document}

\pagestyle{fancy}
\thispagestyle{plain}
\fancypagestyle{plain}{

\fancyhead[C]{\includegraphics[width=18.5cm]{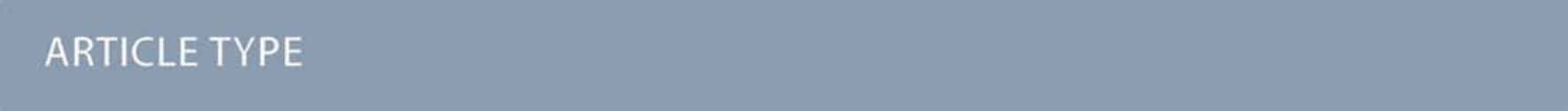}}
\fancyhead[L]{\hspace{0cm}\vspace{1.5cm}\includegraphics[height=30pt]{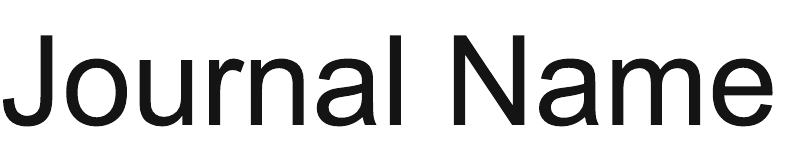}}
\fancyhead[R]{\hspace{0cm}\vspace{1.7cm}\includegraphics[height=55pt]{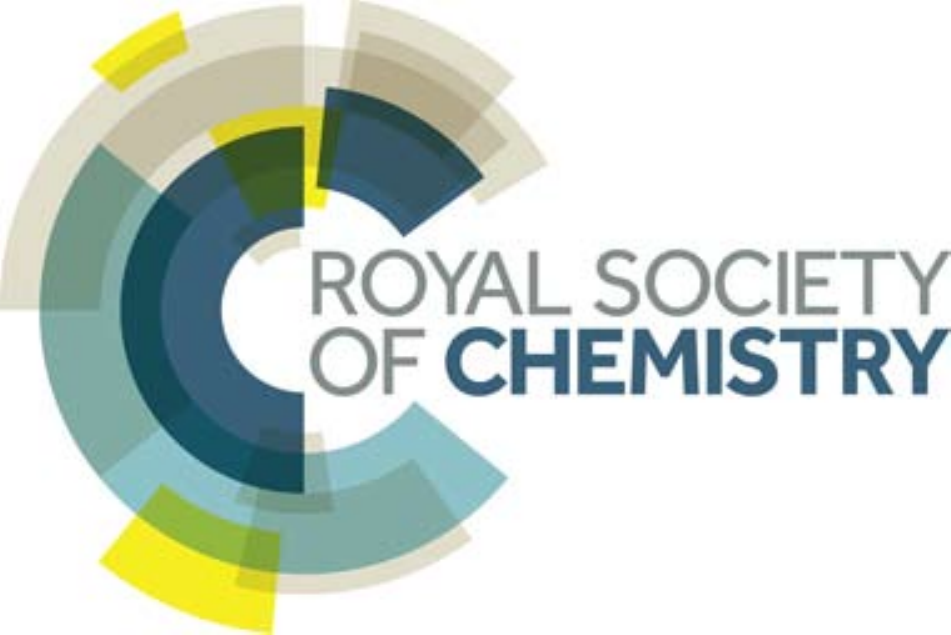}}
\renewcommand{\headrulewidth}{0pt}
}

\makeFNbottom
\makeatletter
\renewcommand\LARGE{\@setfontsize\LARGE{15pt}{17}}
\renewcommand\Large{\@setfontsize\Large{12pt}{14}}
\renewcommand\large{\@setfontsize\large{10pt}{12}}
\renewcommand\footnotesize{\@setfontsize\footnotesize{7pt}{10}}
\makeatother

\renewcommand{\thefootnote}{\fnsymbol{footnote}}
\renewcommand\footnoterule{\vspace*{1pt}%
\color{cream}\hrule width 3.5in height 0.4pt \color{black}\vspace*{5pt}} 
\setcounter{secnumdepth}{5}

\makeatletter 
\renewcommand\@biblabel[1]{#1}            
\renewcommand\@makefntext[1]%
{\noindent\makebox[0pt][r]{\@thefnmark\,}#1}
\makeatother 
\renewcommand{\figurename}{\small{Fig.}~}
\sectionfont{\sffamily\Large}
\subsectionfont{\normalsize}
\subsubsectionfont{\bf}
\setstretch{1.125} 
\setlength{\skip\footins}{0.8cm}
\setlength{\footnotesep}{0.25cm}
\setlength{\jot}{10pt}
\titlespacing*{\section}{0pt}{4pt}{4pt}
\titlespacing*{\subsection}{0pt}{15pt}{1pt}

\fancyfoot{}
\fancyfoot[LO,RE]{\vspace{-7.1pt}\includegraphics[height=9pt]{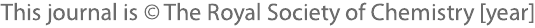}}
\fancyfoot[CO]{\vspace{-7.1pt}\hspace{13.2cm}\includegraphics{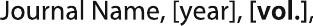}}
\fancyfoot[CE]{\vspace{-7.2pt}\hspace{-14.2cm}\includegraphics{head_foot/RF}}
\fancyfoot[RO]{\footnotesize{\sffamily{1--\pageref{LastPage} ~\textbar  \hspace{2pt}\thepage}}}
\fancyfoot[LE]{\footnotesize{\sffamily{\thepage~\textbar\hspace{3.45cm} 1--\pageref{LastPage}}}}
\fancyhead{}
\renewcommand{\headrulewidth}{0pt} 
\renewcommand{\footrulewidth}{0pt}
\setlength{\arrayrulewidth}{1pt}
\setlength{\columnsep}{6.5mm}
\setlength\bibsep{1pt}

\makeatletter 
\newlength{\figrulesep} 
\setlength{\figrulesep}{0.5\textfloatsep} 

\newcommand{\topfigrule}{\vspace*{-1pt}%
\noindent{\color{cream}\rule[-\figrulesep]{\columnwidth}{1.5pt}} }

\newcommand{\botfigrule}{\vspace*{-2pt}%
\noindent{\color{cream}\rule[\figrulesep]{\columnwidth}{1.5pt}} }

\newcommand{\dblfigrule}{\vspace*{-1pt}%
\noindent{\color{cream}\rule[-\figrulesep]{\textwidth}{1.5pt}} }

\makeatother

\twocolumn[
  \begin{@twocolumnfalse}
\vspace{3cm}
\sffamily
\begin{tabular}{m{4.5cm} p{13.5cm} }

\includegraphics{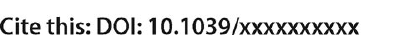} & \noindent\LARGE{\textbf{The effect of curvature on cholesteric doughnuts}} \\
\vspace{0.3cm} & \vspace{0.3cm} \\

 & \noindent\large{Ana Fialho,\textit{$^{a\,\ddag}$} Nelson R. Bernardino,\textit{$^{a\,\P}$} Nuno M. Silvestre,$^{\ast}$\textit{$^{a}$} and Margarida M. Telo da Gama\textit{$^{a}$}} \\

\includegraphics{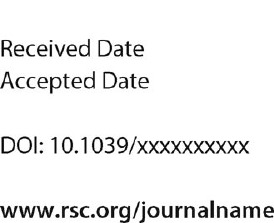} & \noindent\normalsize{ The confinement of liquid crystals inside curved geometries leads to exotic structures, with applications ranging from bio-sensors to optical switches and privacy windows. Here we study how curvature affects the alignment of a cholesteric liquid crystal. We model the system on the mesoscale using the Landau-de Gennes model. Our study was performed in three stages, analysing different curved geometries from cylindrical walls and pores, to toroidal domains, in order to isolate the curvature effects. Our results show that the stresses introduced by the curvature influence the orientation of the liquid crystal molecules, and cause distortions in the natural periodicity of the cholesteric that depend on the radius of curvature and on the commensurability between the pitch and the dimensions of the system. In particular, the cholesteric layers of toroidal droplets exhibit a symmetry breaking not seen in cylindrical pores and that is driven by the additional curvature. } \\

\end{tabular}

 \end{@twocolumnfalse} \vspace{0.6cm}

  ]

\renewcommand*\rmdefault{bch}\normalfont\upshape
\rmfamily
\section*{}
\vspace{-1cm}


\footnotetext{\textit{$^{a}$ Departamento de F{\'\i}sica da Faculdade de Ci\^encias,
Centro de F{\'\i}sica Te\'orica e Computacional,
Universidade de Lisboa, P-1649-003 Lisboa, Portugal
E-mail: nmsilvestre@fc.ul.pt}}


\footnotetext{\ddag Present address: School of Physics and Astronomy, University of Edinburgh, Peter Guthrie Tait Road, Edinburgh EH9 3FD, UK}
\footnotetext{\P Present address: Haitong Bank, S.A., Rua Alexandre Herculano, 38, P-1269-180 Lisboa, Portugal} 



\section{Introduction}

Liquid crystals (LCs) have dominated the display industry for over 50 years and are of standard use in small everyday devices.\cite{Castellano.2005,Hilsum.2010,Choi.2016} Typically, such technology comprises a nematic LC confined between two flat plates, as in monitors,\cite{Scheffer.1984,Tarumi.2002} or encapsulated to cavities dispersed in a polymer matrix (PDLC), as in privacy windows.\cite{Doane.1988,Kitzerow.1994,Drzaic.2006} These rely on the fact that LCs are fluid flexible media extremely sensitive to the confining surfaces and to applied external perturbations.

Nematics are the simplest of the LC phases. They differ from the isotropic phase by exhibiting long range orientational order, which makes it energetically favourable for the nematic molecules to be (on average) uniformly aligned. In the presence of frustration, e.g., imposed by confinement, the nematic induces the nucleation of topological defects (in the orientational field).\citep{Gennes.1993,Chaikin.1995} These are small regions of reduced orientational order that scale with the correlation length and play a major role in wetting,\cite{Patricio.2011} phase transitions,\cite{Chaikin.1995} and in the interaction of colloidal particles dispersed in an LC.\cite{Stark.2001} Their nucleation is particularly important in the physics of bistable LC devices used in a large scale to indicate the price of any produce at your local supermarket.\cite{Xu.2013}

\begin{figure}[t]
\includegraphics[width=0.9\columnwidth]{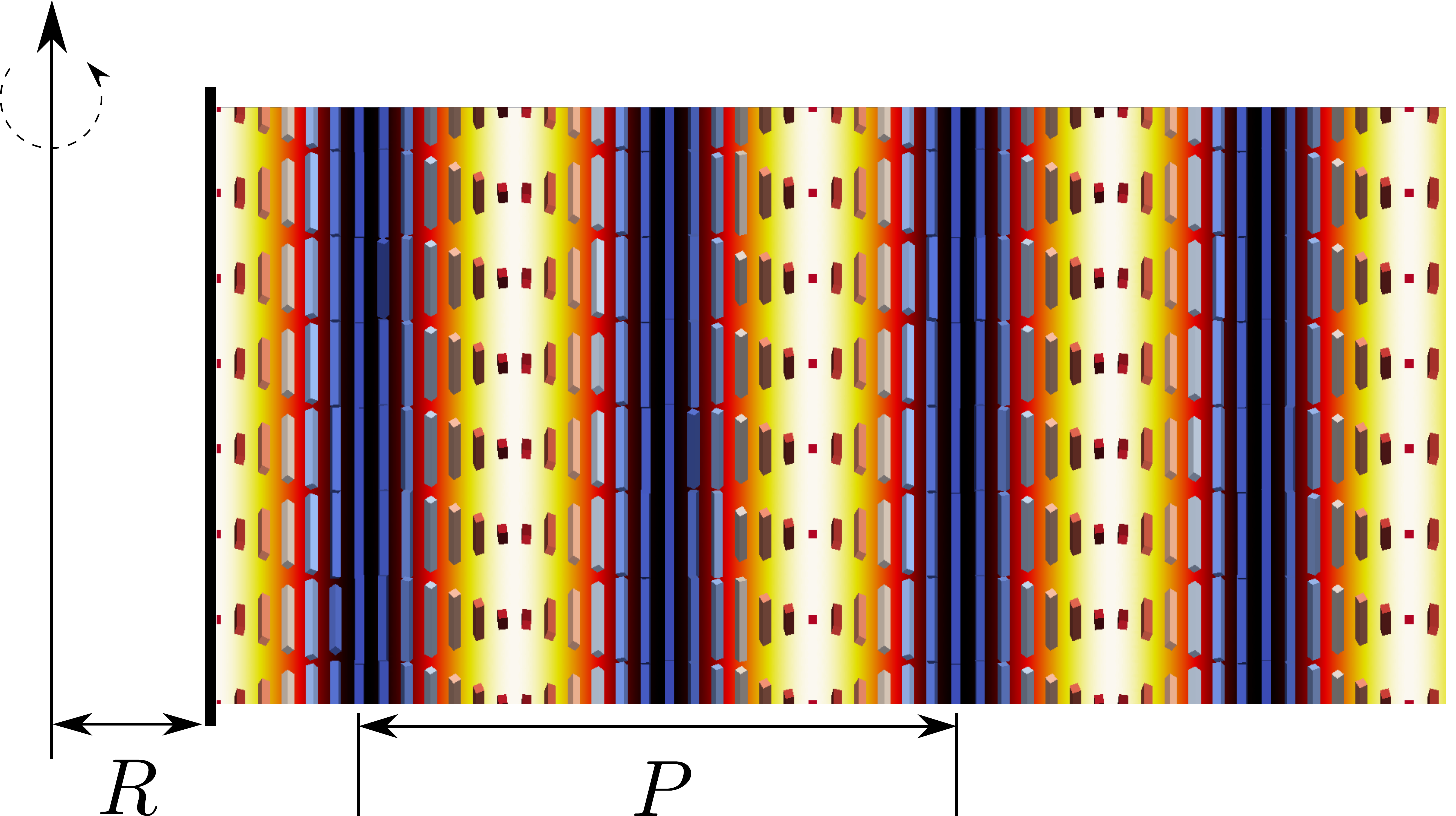}
\caption{Schematic of a cholesteric in contact with a  wall (dark bar on the left) of curvature radius $R$. The pitch $P$ is defined here as the distance needed for the orientational field to rotate by $2\pi$. The configuration presented corresponds to the limit of a flat surface ($R\rightarrow\infty$). The color indicates the off-plane (angular, $n_\theta$) component of the orientational field. White: the director points out of the plane; Black: the director is in-plane.}
\label{fig:schematic}
\end{figure}

\begin{figure}[t]
\includegraphics[width=1.0\columnwidth]{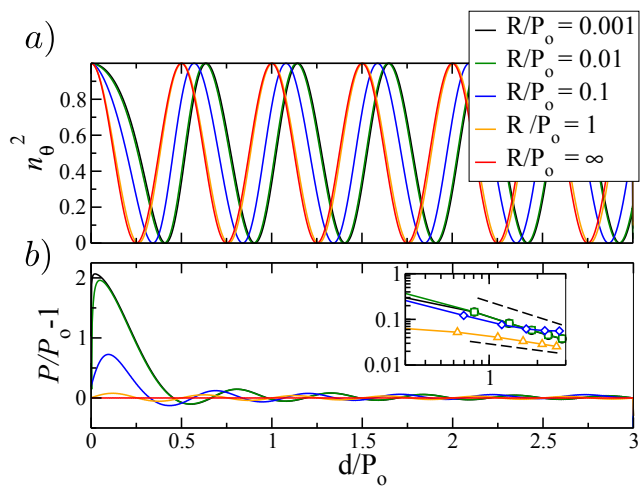}
\caption{(a) Square of the off-plane (angular) component of the director field $n_\theta^2$ and  (b) cholesteric pitch $P$ as functions of the distance to the wall $d$ with curvature radii $R/P_o=0.001,\,0.01,\,0.1,\,1,\,\infty$. $P_o$ is the natural pitch of the system. In the inset: maxima of $P$ as a function of the distance. The dashed lines correspond to $P/P_o-1\sim (d/P_o)^{-\alpha}$ with $\alpha=0.01,\,0.05$ and are represented for comparison.}
\label{fig:PvsR}
\end{figure}

Also with a wide range of applications are cholesteric LCs. From the macroscopic point of view, they differ from nematics by the fact that they exhibit spontaneous twist that, on the microscopic level, is the result of being composed of chiral molecules.\citep{Gennes.1993,Chaikin.1995} To the general public they are probably better known for their application as mood rings and thermometer strips, \cite{White.1999,Jiang.2002} and eWriters.\cite{Montbach.2016} A key feature of cholesteric LCs important for applications is the ability to control the reflected light by manipulating its spontaneous pitch, either by the application of external fields or by changing the temperature of the LC. This seems to be completely understood from the point of view of cholesterics confined to flat surfaces. However, as the display technology moves towards softer hardware it is of crucial importance to understand what is the effect of curvature on the cholesteric pitch. 

The development of new soft-lithographic techniques made possible the controlled production of new PDLC matrices with droplets of prescribed shapes and non-trivial topologies.\cite{Campbell.2014} Of particular interest are toroidal shapes, for which the Poincar\'e-Hopf theorem dictates that the net charge of the bulk defects is $\sum_i{q_i}=0$, in the absence of surface defects. For a torus on the micrometer scale defects may appear in pairs of opposite charge.\cite{Campbell.2014} However, as the system size increases the presence of topological defects becomes energetically unstable, and for a torus on the millimetre scale, the LC texture is completely free from topological defects.\cite{Pairam.2013} However, in some toroidal systems the liquid crystal texture can contain defects. This happens when two regions on the confining surface have different boundary conditions. \cite{Batista.2015} Such is the case of liquid crystal droplets adsorbed at micrometer fibers.\cite{Geng.2012,Geng.2013}

In this work we are interested in understanding what is the role of curvature on the texture of cholesteric liquid crystals confined to toroidal domains. This manuscript is organized in the following manner. In the next section we describe how we model the liquid crystal and briefly describe our numerical methods. To achieve our aim we have divided our study into three stages, which are presented in Sec. \ref{sec:results}.
 First, we consider a cholesteric liquid crystal in contact with a cylindrical substrate and analyse how is the cholesteric affect by decreasing the radius (increasing the curvature) of the cylindrical wall. In particular we will show how adding curvature changes the local periodicity of the cholesteric. Second, we consider that the cholesteric is confined to a cylindrical pore and revisit some of the results obtained by Ambrozic and Zumer. \cite{Ambrozic.1996,Ambrozic.1999} We will show that when the periodicity of the cholesteric is non-commensurate with the size of the cylinder that liquid crystal is no longer symmetric. Finally, we consider the case of a toroidal domain and analyse how the curvature of the confining surface affects the cholesteric texture.
We summarize our results in Sec. \ref{sec:conclusions}.

\section{Model}
We use the Landau-de Gennes model to describe the cholesteric LC.\cite{Gennes.1993} The order parameter is a traceless, symmetric 2nd-rank tensor $\Qvec$ that carries information on the local average molecular orientation, the director field $\nvec$, and on the degree of orientational order $S$; for uniaxial nematics it takes the form $Q_{\alpha\beta}=S\left(3n_\alpha n_\beta-\delta_{\alpha\beta}\right)/2$, where $\delta_{\alpha\beta}$ is the Kronecker delta. The free energy is written in invariant terms of $\Qvec$ and its derivatives $\partial\Qvec$, $F_{LdG}(\Qvec,\partial\Qvec)=\int_\Omega{dV\,\left(f_b(\Qvec)+f_e(\partial\Qvec)+f_{q_o}(\partial\Qvec)\right)}+\int_{\partial\Omega}{ds\,f_s(\Qvec)}$, with free energy densities as \cite{Bernardino.2014}

\begin{eqnarray}
f_b&=& \frac{2}{3} \tau \Tr{\Qvec^2}-\frac{8}{3}\Tr{\Qvec^3}+\frac{4}{9} \left(\Tr{\Qvec^2}\right)^2, \\
f_e&=&\frac{1}{3}\left(\partial_k Q_{ij}\right)^2 ,\\
f_{q_o}&=&\frac{4q_o}{3}Q_{il}\epsilon_{ijk}\partial_j Q_{kl}.
\end{eqnarray}
Where $\Tr$ is the trace operator. The bulk free energy $f_b$ describes the phase transition between the isotropic and the LC phases, the elastic contribution $f_e$ is the same for nematics while $f_{q_o}$ accounts for the spontaneous twist deformations observed in cholesteric phases. Here we adopt a dimensionless model and thus the free energy depends only on two bulk parameters $\tau \,\mbox{and }q_o$, and length is measured in units of the correlation length $\xi$; as a reference, for a nematic such as the 5CB at room temperature $\xi\simeq 15$ nm. $\tau$ is a reduced temperature. Its value determines the equilibrium bulk phase. $q_o=2\pi/P_o$ is the inverse of the cholesteric pitch $P_o$. In this simple version of the Landau-de Gennes theory we allow for splay, twist, and bend deformations to have, in the limit of $P_o\rightarrow \infty$, the same energy cost, also known as the one-elastic constant approximation.

To account for the presence of confining substrates, a surface free energy can be introduced. We restrict our study to substrates that impose parallel alignment of the nematic molecules. In particular, in the full three dimensional problem (see below), we consider the Fournier-Galatola planar degenerate surface potential \cite{Fournier.2005}

\begin{equation}
f_s = \frac{W}{2}\left(\Tr{\left(\tilde{\Qvec}-\tilde{\Qvec}^\perp\right)}^2
+\left(\Tr{\tilde{\Qvec}}-\frac{3}{2}S^2_s\right)^2\right),
\end{equation}
where $W$ is the reduced surface anchoring constant, $\tilde{Q}_{\alpha\beta}=Q_{\alpha\beta}+S_s\delta_{\alpha\beta}/2$ and $\tilde{\Qvec}^\perp=\Pvec^T\tilde{\Qvec}\Pvec$, with the projection matrix $P_{\alpha\beta}=\delta_{\alpha\beta}-\nu_\alpha\nu_\beta$; $\vec{\nu}$ is the surface normal. The first term penalizes deviations of the director field from any (local) parallel orientation on the substrate. The second term forces the degree of orientational order to coincide with $S_s$, the preferred degree of orientational order at the surface.

In this work we address the effect of curvature on the cholesteric pitch. To that end we are interested in measuring the local pitch $P(\rvec)$ that characterizes the local twist of the director field $\nvec$. This is done by evaluating the twist parameter \cite{Kilian.1995,Copar.2013}
$S_{tw}=\varepsilon_{ijk}Q_{il}\partial_j Q_{lk}$, which is inversely proportional to the pitch
\begin{equation}
\frac{P(\rvec)}{P_0} = -\frac{9}{4} \frac{S^2}{S_{tw}}q_0.
\label{eq:pitch}
\end{equation}
The sign is related to the direction of rotation of the twist deformation. $S_{tw}$ is negative if the rotation is anti-clockwise and positive if clockwise.

This study was performed in different steps. We first considered cholesteric systems with either translational or cylindrical symmetry. This allows to simplify the problem to two dimensions (2D). Our 2D studies were performed with COMSOL 3.5a, \cite{COMSOL.2005} which uses finite elements techniques to solve the Euler-Lagrange equations that follow from the minimization of the Landau-de Gennes free energy. Details on this method can be found in \cite{Fialho.2015}. Finally, we considered the full three dimensional problem of toroidal droplets and minimized the free energy with finite element methods with adaptive meshing, described in detail in \cite{Tasinkevych.2012}. In both 2D and 3D studies our meshes are refined to insure a numerical precision $<1\%$ for the Landau-de Gennes free energy.

\section{Results}
\label{sec:results}
\subsection{Cholesteric near a flat surface}

\begin{figure}[t]
\includegraphics[width=0.8\columnwidth]{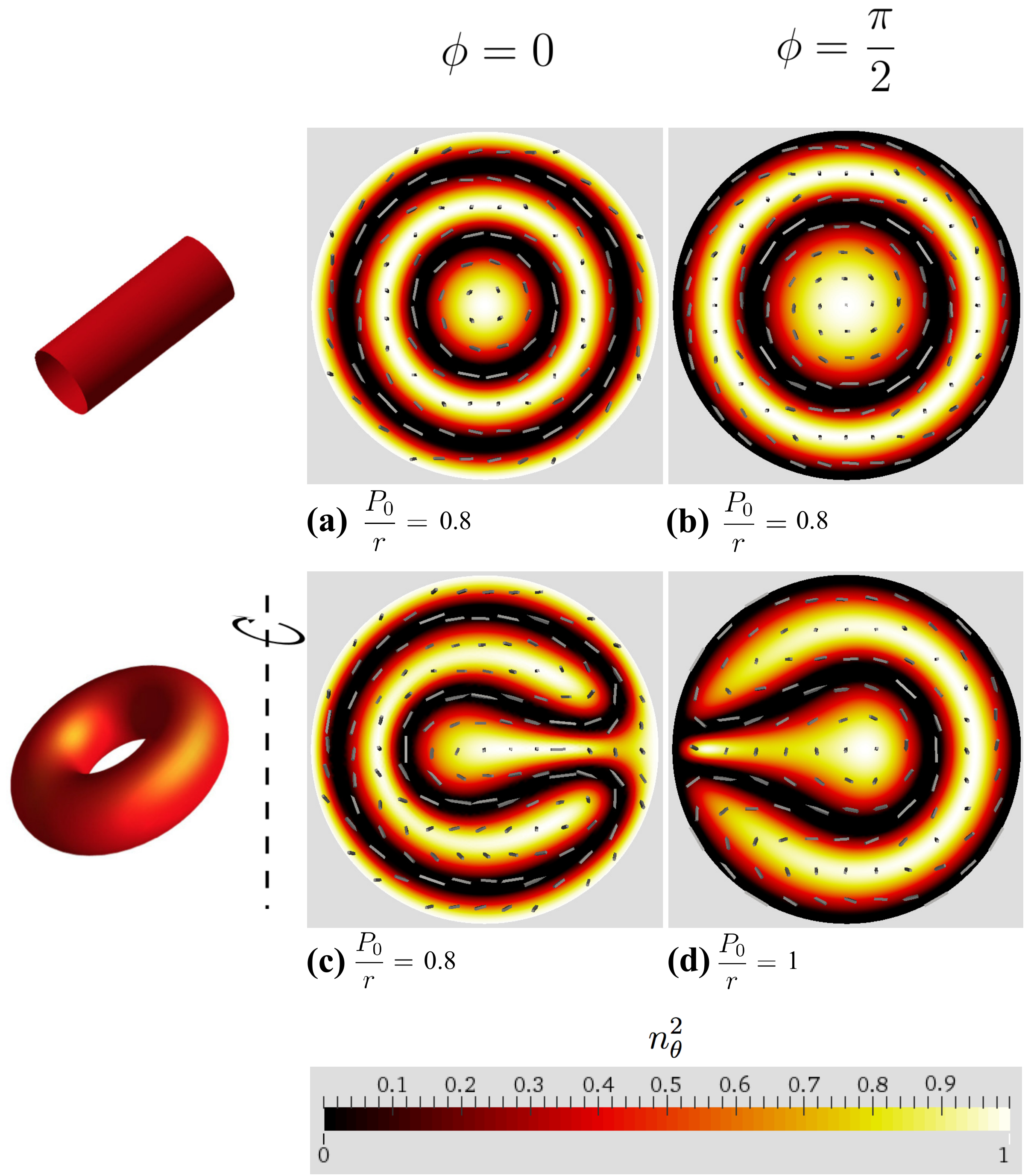}
\caption[Comparison between the cholesteric configurations for the no curvature limit and for a curvature radius $R=50\xi$.]{Comparison between the cholesteric configurations for the no curvature limit ((a) and (b)) and for a curvature radius $R=50\xi$ ((c) and (d)). Systems (a) and (c) have boundary conditions with $\phi=0$ and systems (b) and (d) have $\phi=\frac{\pi}{2}$. The color represents $n_{\theta}^2$, the square of the component of the director along $\textbf{e}_{\theta}$. The grey bars represent the director field.}
\label{fig:cylTor}
\end{figure}

When a cholesteric phase is in contact with a flat substrate the anchoring on the surface forces the cholesteric LC to assume a preferred configuration. For example, if the surface induces planar anchoring, the cholesteric orients its twist axis in such a way that the cholesteric layers are aligned parallel to the substrate. If by any means the cholesteric were to align its twist axis parallel to the substrate, the orientational frustration imposed by the surface anchoring would induce the nucleation of evenly spaced defects at the substrate, thus increasing the free energy of the system. 
Such configuration is only preferred if the anchoring on the substrate is homeotropic as in that case there is no other  configuration with lower energy cost.\cite{Silvestre.2016} Here, nonetheless, we focus on the case of a surface with planar anchoring.

In Fig. \ref{fig:schematic} we show the preferred configuration of a cholesteric LC near a flat wall (dark bar on the left) with planar anchoring. The cholesteric twists in the direction perpendicular to the substrate and forms a layer-like system. The phase is characterised by a twist pitch $P$, which is constant throughout the entire system and coincides with the natural cholesteric pitch $P_o$ that is an input to the model, and describes the undulation of the director field, shown in Fig.\ref{fig:PvsR}a for $R/P_o=\infty$.

\subsection{Adding curvature: from flat to cylindrical surfaces}
\label{sec:cylinder}

\begin{figure}[t]
\includegraphics[width=0.9\columnwidth]{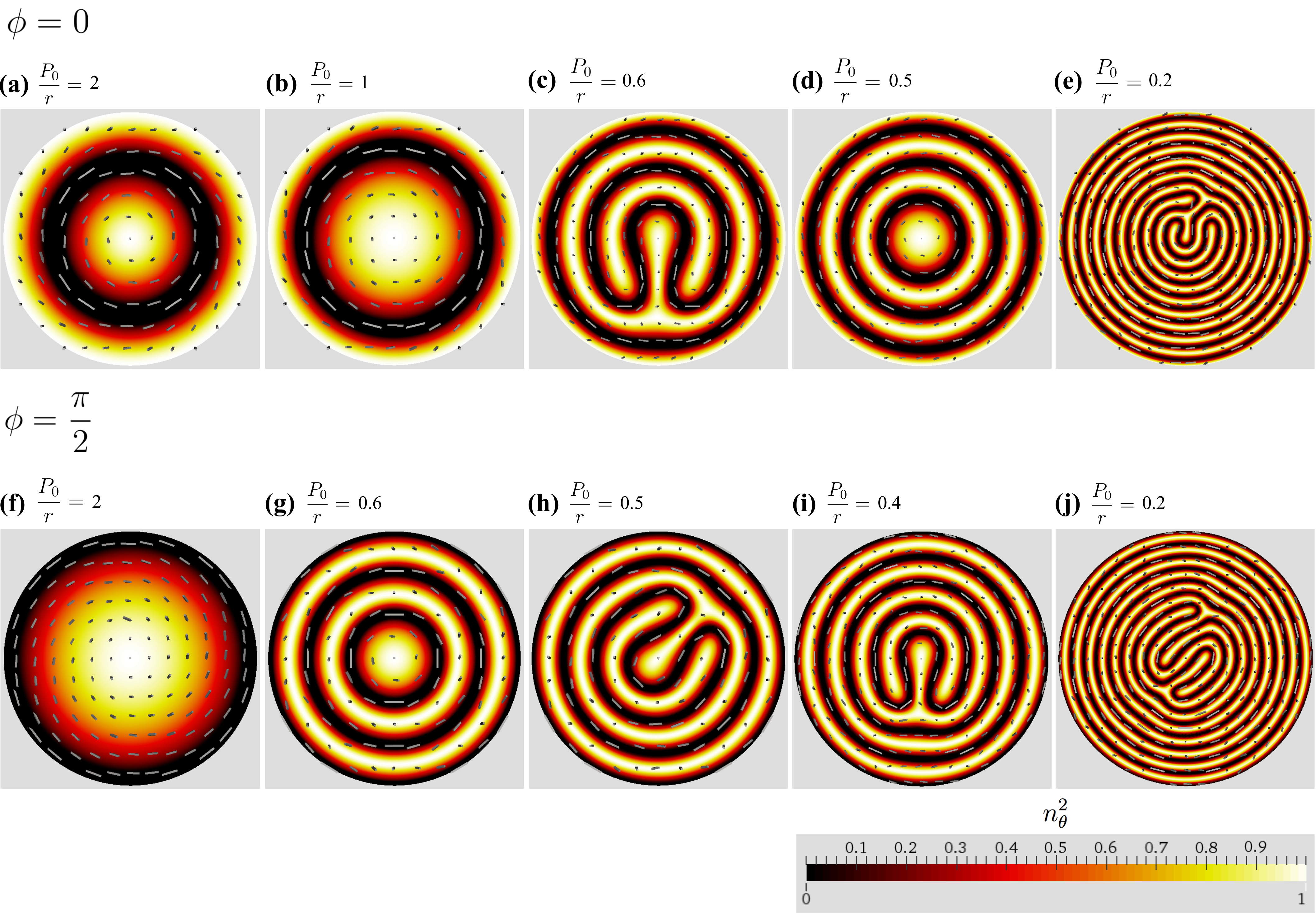}
\caption[Equilibrium configurations for a cholesteric liquid crystal inside an infinite cylinder with radius $r=500\xi$.]{Equilibrium configurations for a cholesteric liquid crystal inside an infinite cylinder with radius $r=500\xi$. (a) to (e): fixed boundary conditions with director parallel to the long axis of the cylinder ($z$-axis). (f) to (j) fixed boundary conditions with director tangent to the limiting circumference, on planes of constant $z$. The color represents $n_{\theta}^2$, the square of the component of the director along $\textbf{e}_{\theta}$. The grey bars represent the director field.}
\label{fig:cylinder}
\end{figure}

To address the effect of curved surfaces we consider that the wall is the surface of a cylinder of radius $R$, as defined in Fig.\ref{fig:schematic}. In this case the flat surface corresponds to the limit $R\rightarrow \infty$ for which the cholesteric 'layers' are equally spaced by $P_o/2$. In Fig.\ref{fig:PvsR}a we show how the square of the off-plane director component $n_\theta^2$ varies with the distance to the wall $d$ for different radii $R/P_o=0.001,\,0.01,\,0.1,\,1,\,\infty$. $n_\theta^2$ is an harmonic function of the distance varying from $0$ to $1$. Clearly, if the radius of curvature is comparable to the natural pitch $P_o$ then $n_\theta^2$ is  very similar to the one obtain for the flat wall ($R/P_o=\infty$). However, if the radius of curvature  is much smaller than the pitch Fig.\ref{fig:PvsR}a shows that $n_\theta^2$ is shifted significantly and its maxima and minima occur farther away from the wall when compared with the flat case. This indicates that as the curvature increases ($R$ decreases) the actual pitch of the cholesteric deviates from its natural value $P_o$.

Figure \ref{fig:PvsR}b shows how the local pitch $P$ (defined in Eq.\ref{eq:pitch}) varies with the distance to the wall $d$ for different radii $R/P_o=0.001,\,0.01,\,0.1,\,1,\,\infty$. For the flat wall the pitch $P=P_o$ is constant everywhere, as already discussed. As the radius of the cylinder is decreased the local pitch $P$ undulates close to the natural pitch value and converges to $P_o$ far from the wall $d\rightarrow\infty$. By increasing the curvature of the wall the undulations of $P$ are more pronounced close to the substrate, but rapidly decay as $d$ increases. However, for distances $d > 0.75P_o$ the decay is very slow as can be shown in the inset of Fig.\ref{fig:PvsR}b. The points indicate the maxima of $P$ and the dashed lines correspond to $P/P_o-1\sim (d/P_o)^{-\alpha}$ where $\alpha=0.01$ (bottom line) and $\alpha=0.05$ (upper line). 

The undulating behaviour seen here is related to the compression $P<P_o$ and dilation $P>P_o$ of the cholesteric, and it comes from the energetic cost associated with a bend deformation induced by the cylindrical symmetry, allied with the cholesteric twist. 
For example, consider the cholesteric layer in which the off-plane component reaches zero. In this situation, the bending energy associated with bending such 'layer' is nearly zero as, in this region, $\partial_\theta n_\theta\sim 0$. However, the bending energy is large for the layers in which $n_\theta^2=1$. This indicates that the system can compromise between dilating the layers with $n_\theta=0$ and compressing those with $n_\theta^2=1$.

\subsection{Cholesteric confined to cylinders}

We now consider that the cholesteric is confined to a cylindrical tube of radius $r$. Such system was previously studied by Ambrozic and {\v Z}umer.\cite{Ambrozic.1996,Ambrozic.1999} Here, we restrict our study to configurations that have translational symmetry, and consider two different orientations on the surface of the cylinder, shown in Fig.\ref{fig:cylTor}a and b for $r/P_o=1.25$ : (i) $\nvec$ is along the symmetry axis of the cylinder and (ii) the director is in-plane and parallel to the surface of the cylinder. In both cases the cholesteric twists radially forming an onion-like configuration. Also in this situation the layers compress or dilate depending on if the director field is, respectively, in-plane (dark regions) or off-plane (bright regions). However, in this situation the configuration of the cholesteric also depends on how the radius of the cylinder commensurates with $P_o$.

Figure \ref{fig:cylinder} shows a cholesteric LC inside a long cylinder of radius $r=500\xi$ for different pitch $P_o$ values and the same two types of boundary conditions shown in Fig.\ref{fig:cylTor}. In the configurations found, the cholesteric at the center of the cylinder is always pointing off-plane. This means that when at the boundary the director is also off-plane, the cholesteric will try to perform (radially) at least a half turn (see Fig.\ref{fig:cylinder}a). However, if the $\nvec$ at the surface of the cylinder is in-plane, the minimum rotation that the cholesteric can perform is $1/4$ of a turn (see Fig.\ref{fig:cylinder}f). 

For both cases, whenever the pitch commensurates with the radius of the cylinder the cholesteric displays an onion-like configuration with the cholesteric layers arranging themselves as concentric cylinders. However, when the pitch is non-commensurate with $r$ the layers are disrupted and assume a configuration that resembles that of finger prints. In this situation, the stress that comes from the compression and dilation of the cholesteric layers to maintain the onion-like configuration is high and the system prefers to split some cholesteric layers, particularly the ones closer to the center as it is there that the bending energy is higher.

\subsection{Full confinement: from a cylinder to a torus}

\begin{figure}[t]	
\includegraphics[width=1.0\columnwidth]{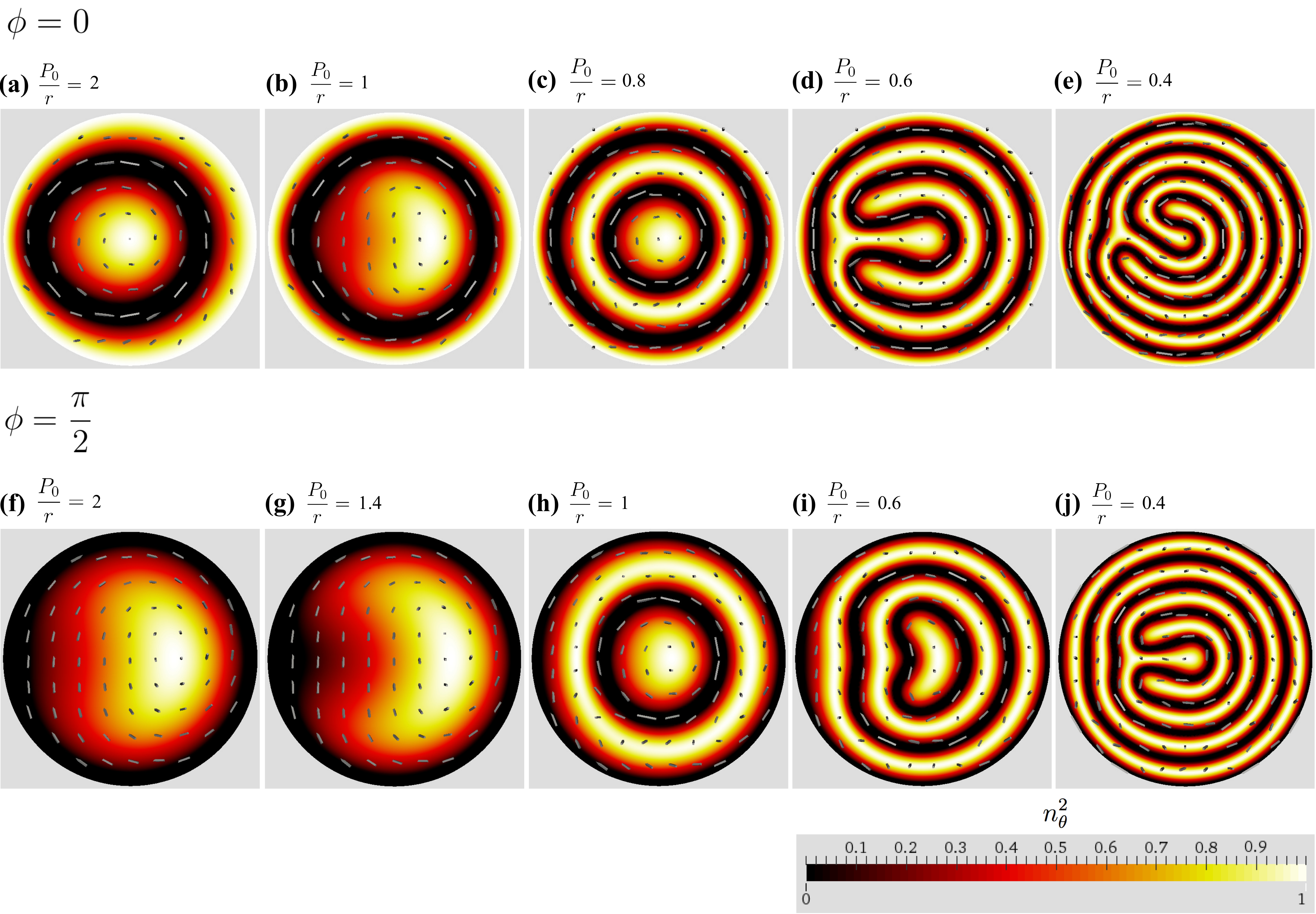}
\caption[Cholesteric configurations for a system with radius of curvature $R=50\xi$.]{Cholesteric configurations for a system with radius of curvature $R=50\xi$. (a) to (e) correspond to boundary conditions with $\phi=0$, and (f) to (j) have boundary conditions with $\phi=\frac{\pi}{2}$. The color represents $n_{\theta}^2$, the square of the component of the director along $\textbf{e}_{\theta}$. The grey bars represent the director field.}
\label{torus2d}
\end{figure}

\begin{figure}[t]
\includegraphics[width=1.0\columnwidth]{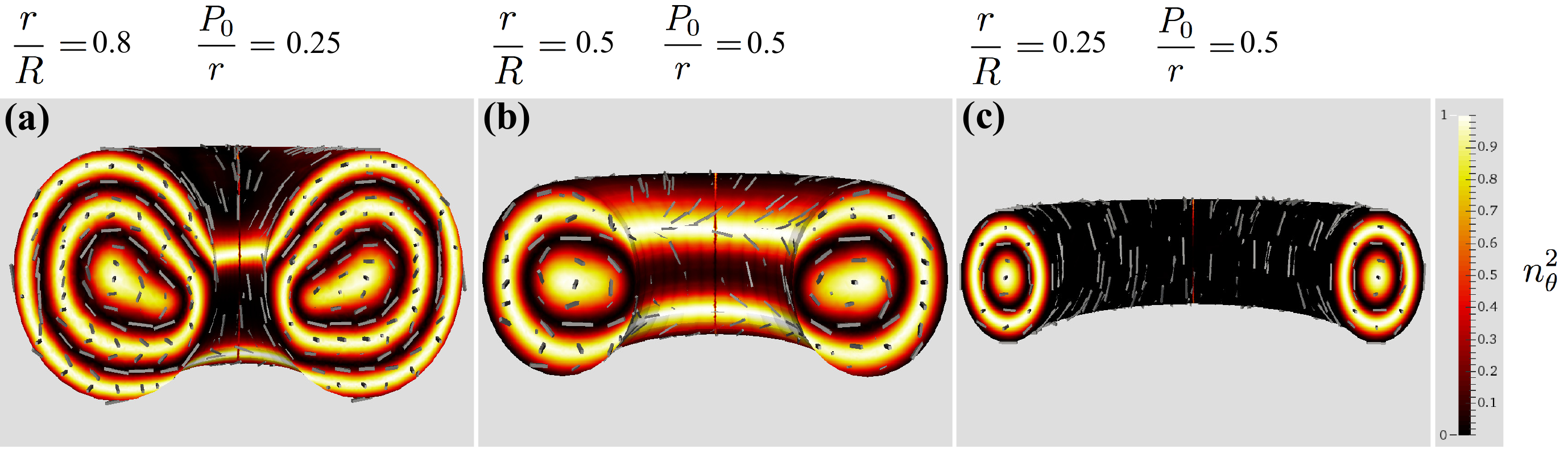}
\caption[Configurations of a cholesteric inside a torus, with multiple layers.]{Configurations of a cholesteric inside a torus, with multiple layers. (a) $\frac{r}{R} = 0.8 , \frac{P_0}{r} = 0.25$, (b) $\frac{r}{R} = 0.5 , \frac{P_0}{r} = 0.5$, (c) $\frac{r}{R} = 0.25 , \frac{P_0}{r} = 0.5$. The color scale refers to $n_{\theta}^2$ and the grey bars represent the director field.}
\label{fig:torus}
\end{figure}

Confining the cholestic LC inside a cylindrical pore clearly affects the local pitch, not only due to the curvature of the confining surface but also due to the competing length scales, the pitch $P$ and the radius of the cylinder $r$.

Within good approximation a torus (doughnut) of infinite principal radius, $R\rightarrow\infty$, corresponds to a long cylinder. This means that as the principal radius of the torus is decreased the cholesteric layers confined inside the torus should respond accordingly to the additional curvature as depicted in Figs.\ref{fig:cylTor}c and \ref{fig:cylTor}d.

In Fig.\ref{torus2d} we show the cholesteric layers inside a torus of principal radius $R=50\xi$ and cross-section $r=500\xi$ for different pitch values $P_o$. Again, we consider two types of anchoring orientations at the surface of the torus: \ref{torus2d}a -- \ref{torus2d}e) in-plane $\phi=0$, and \ref{torus2d}f -- \ref{torus2d}j) off-plane $\phi=\pi/2$. The (cylindrical) symmetry axis is located on the left. As in Section \ref{sec:cylinder}, when the director is parallel to the symmetry axis there is no additional bending energy associated to extending that particular region, and whenever the director if perpendicular to the symmetry axis (off-plane) it is preferable to contract that region to reduce bend deformations. As a result, even for the case of concentric layers in the $R\rightarrow \infty$ limit the layers are deformed.

As in the case of a cylindrical pore, when the cross-section of the toroidal droplet is non-commensurate with the pitch, the central layers break and allow the outer layers to extend towards the center of the cross-section. Here the additional ingredient is that this extension of the outer layers mainly happens along the equatorial plane of the torus, which indicates that there is a constraint on this extensional direction  that is set by the additional curvature.

Finally, we have considered the full 3D problem. Figure \ref{fig:torus} shows the configuration of cholesterics LC confined to toroidal droplets with planar degenerate anchoring, i.e., here the LC molecules are allowed to align in any direction parallel to the surface of the torus. As a result, the cholesteric is now able to relief some of the built-in stress due to the curved surfaces that confines it by assuming different orientations on the same cross-section. In this way, the cholesteric now exhibits broken layers that penetrate the surface. We note that on the equatorial plane, in the inner most region of the surface of the torus, where the curvature is highest, the director field is parallel to the symmetry axis, thus preventing to increase the free energy of the system through bend deformation.

\section{Conclusions}
\label{sec:conclusions}

In this manuscript we have considered the effect of curvature on the pitch of a cholesteric liquid crystal.
We started by considering a cholesteric LC in contact with a cylindrical wall of a given radius. We found that, because of the curvature, the cholesteric is subjected to additional bend distortions that influence the orientational field. As a result, the cholesteric contracts the regions where this bend deformation is higher and dilates the ones where it is lower, thus changing locally the value of the pitch $P$. These distortions were found to be larger for higher curvatures (thinner cylinders) but rapidly decaying with the distance to the wall.

When the cholesteric is insider a cylindrical cavity the distortion effect due to the curvature of the confining wall is again present. Typically, the cholesteric assumes an onion-like configuration with the cholesteric 'layers' adopting the form of concentric cylinders. However, if the radius of the cylinder was non-commensurate with the cholesteric pitch, due to the stress coming from the bend deformations, the inner most layers split and the cholesteric assumes a configuration resembling finger prints.

We then considered the cholesteric to be confined in a toroidal droplet, with cylindrical symmetry around its symmetry axis. For an infinitely large torus the cholesteric adopts the same configurations as in the cylindrical confinement case with a given cross-sectional radius. As the torus is made smaller there is a shift in the position of the cholesteric layers that appears as a result of the additional curvature. The effect is more pronounced for smaller radii.

Finally, we considered the full three-dimensional toroidal droplet and considered planar degenerate anchoring on the surface of the torus. We corroborate the results found for systems with cylindrical symmetry. Particularly, we found that the distortion of the natural pitch depends heavily on the  curvature of the droplet. However, because the system is less constrained, particularly at the confining surface, we observe an enhancement of the effect seen for non-commensurable length scales that induce the splitting of the cholesteric layers.

Our main conclusion is that the curvature, and in particular that of the toroidal droplets, affects the local orientation of the cholesteric LC distorting the periodicity of the spontaneous twist deformation. For the case of confining surfaces, this results is a symmetry breaking of the cholesteric layers that depends not only on the curvature of the bounding surface but also on the commensurability between the pitch and the geometrical parameters.

\section*{Acknowledgements}

We acknowledge the financial support of the Portuguese Foundation for Science and Technology (FCT) under the contracts numbers UID/FIS/00618/2013 and EXCEL/FIS-NAN/0083/2012.

\balance

\bibliography{rsc}
\bibliographystyle{rsc}

\end{document}